\documentstyle[12pt,german,epsf,rotate]{article}

\advance\textheight by \topskip
\renewcommand{\baselinestretch}{1.2}
\renewcommand{\thefootnote}{\fnsymbol{footnote}}
\newcommand{\beq}{\begin{equation}}      
\newcommand{\eeq}{\end{equation}}
\newcommand{\bea}{\begin{eqnarray}}
\newcommand{\eea}{\end{eqnarray}}

\renewcommand{\bar}[1]{\overline{#1}}

\newcounter{hilf}

\begin{document}
%%%%%%%%%%%%%%%%%%%%%%%%%%%%%%%%%%%%%%%%%%%%%%%%%%%%%%%%%%%%%%%%%%%%%%%%%
%%%%                         Titlepage                               %%%%
%%%%%%%%%%%%%%%%%%%%%%%%%%%%%%%%%%%%%%%%%%%%%%%%%%%%%%%%%%%%%%%%%%%%%%%%%

\begin{titlepage}
  \renewcommand{\baselinestretch}{1}
  \renewcommand{\thefootnote}{\alph{footnote}}

  \thispagestyle{empty}

   {\bf \hfill                                       TUM--HEP--251/96}

\vspace*{-0.3cm}
   {\bf \hfill\hfill                                       July 1996} 

\vspace*{0.5cm} {\Large\bf
  \begin{center} Propagating Higgs Boundstates in SUSY 
  \end{center}}

\vspace*{1cm}  {\large\sc 
\begin{center} 
    Richard Dawid\footnote{\makebox[1.cm]{Email:}
    Richard.Dawid@Physik.TU-Muenchen.DE
    \\ \hspace*{0.55cm} address after August 1996: \\ 
    \hspace*{0.55cm} Institut f\"ur 
    Theoretische Physik, Universit\"at Wien, 
    \\ \hspace*{0.55cm}  Boltzmanng. 5, 1090 Wien, Austria
    \\ \hspace*{0.55cm}  
    \makebox[1.cm]{Email:} Richard.Dawid@ariel.pap.univie.ac.at}
    and Serguei Reznov\footnote{\makebox[1.cm]{Email:}
    Serguei.Reznov@Physik.TU-Muenchen.DE}
               
 \end{center}}
\begin{center}
    Institut f\"ur Theoretische 
    Physik, Technische Universit\"at M\"unchen,          \\
    James--Franck--Stra\ss e, D--85747 Garching, Germany
 \end{center}
\vspace*{2.3cm}

                     {\Large \bf \begin{center} Abstract \end{center} } 

The possibilities for an underlying theory behind an effective 
supersymmetric Nambu--Jona-Lasinio like 
model of electroweak symmetry 
breaking with propagating Higgs boundstates (e.g. supersymmetric
topcondensation) are investigated. 
The concept of a renormalizable underlying theory turns out not to be
appropriate. We argue that the new theory should come from the Planck scale 
and be connected to supergravity. Such a model can be constructed but
necessarily implies a nonlinear definition of the full Lagrangian.

\renewcommand{\baselinestretch}{1.2}
\end{titlepage}

\newpage
\renewcommand{\thefootnote}{\arabic{footnote}}
\setcounter{footnote}{0}

%%%%%%%%%%%%%%%%%%%%%%%%%%%%%%%%%%%%%%%%%%%%%%%%%%%%%%%%%%%%%%%%%%%%%%%%%
%%%%                         Introduction                            %%%%
%%%%%%%%%%%%%%%%%%%%%%%%%%%%%%%%%%%%%%%%%%%%%%%%%%%%%%%%%%%%%%%%%%%%%%%%%

%\section{Introduction}

%%%%%%%   BHL  %%%%%%%%%%%
Whether the electroweak symmetry is spontaneously broken by fundamental scalars
or by a dynamical mechanism still remains an open question. In case of
dynamical symmetry breaking  two alternatives can be distinguished:
Models with no propagating scalars at all like technicolour \cite{TC} 
and models 
which replace the fundamental Higgs by a propagating bound state, 
like e.g.top condensation \cite{BHL}. The latter approach which involves 
propagating scalar boundstates is based on the
separation of the boundstate mass scale and the scale where this boundstate
breaks up. While this separation can be achieved by an albeit undesirable
fine tuning of the four--fermion coupling in standard model(SM)--like 
top condensation,
it turns out to become a major complication in SUSY extensions due a
basically very  attractive SUSY property:
The cancellation of quadratic divergences. The question
of our paper is whether there exists an underlying theory of supersymmetric
top  condensation which is able to accomplish the required 
separation of scales. We will first
give a very short introduction to top condensation and its
supersymmetric extension. Then we will investigate the approach
of introducing heavy particle exchange in an underlying theory to
explain the 4--fermion terms and we will
see that it fails for several reasons. Next we present a new
approach in connection with supergravity which, as far as we can see,
is the only possible solution in the framework of established 
theoretical concepts.

%%%%%%%%%%%%%%%%%%%%%%%%%%%%%%%%%%%%%%%%%%%%%%%%%%%%%%%%%%%%%%%%%%%%%%%%%%

%\section{Top Condensation and its Supersymmetric Extension}

%%%%%%%%%%%%%%%%%%%%%%%%%%%%%%%%%%%%%%%%%%%%%%%%%%%%%%%%%%%%%%%%%%%%%%%%%%%%%

Before dealing with the supersymmetric case let us shortly reconsider 
the mechanism of isospin breaking in non--supersymmetric top condensation.
A new strong
four--fermion interaction is introduced which induces spontaneous 
electroweak symmetry breaking by a self consistent Nambu--Jona-Lasinio
gap equation and generates a composite Higgs
sector. The Lagrangian contains the usual covariant kinetic terms
for all gauge fields, quarks and leptons and the new interaction term
involving left--handed quark doublets $q={t_L \choose b_L}$ and right--handed singlets $t$:

\beq 
{\cal L}_{} = {\cal L}_{{kin}_{cov}} + G\bar{q}t_R\bar{t}q
\label{Lbhl}
\eeq 

For $G> G_{cr} = 8\pi^2/N_c\Lambda^2$ the gap equation 
has an energetically favoured non--trivial solution for a top--mass $m_t>0$ 
and a top condensate emerges. The high energy cutoff 
$\Lambda$ corresponds to new physics which is resolved at this scale, 
e.g. a heavy 
gauge boson exchange. In other words the coupling becomes non--local at this
scale and the scalar bound state breaks up. As there is one free parameter 
less than in the SM,
this theory predicts a certain ratio between the top quark mass 
$m_{t}$ and the W--mass $m_W$ in slight dependence of $\Lambda$. The 
resulting top--mass, becoming slightly smaller with increasing  $\Lambda$
still has an unrealistic value of about 220~GeV for $\Lambda$ at the 
Planck scale. Moreover to achieve the large scale separation between 
$\Lambda$ and $m_t$ one has to fine--tune the four fermion coupling $G$ towards
the critical coupling $G_{cr}$.

A supersymmetric version of a Nambu--Jona-Lasinio Model \cite{BE}, 
formulated in the framework of top condensation in \cite{SBHL}, 
appears to be able to 
solve both problems mentioned above: A lower top mass is natural first due to
a lower value of the quasi fixed point of $g_t$ in the minimal
supersymmetric standard model (MSSM) and second because
of the additional Higgs doublet which allows to lower the top mass by the
factor $sin(\beta)$, the ratio between the values of the 
vacuum expectation values (VEVs) for the two Higgs fields. 
The hierarchy problem which appears as the fine--tuning
problem of $G$ in top condensation is alleviated in supersymmetry 
because of the cancellation of quadratic divergences.

The supersymmetric extension of (\ref{Lbhl}) is

\bea
{\cal L}={\cal L}_{YM}&+ &\int d^{2}\theta d^{2}\bar{\theta}
(\bar{Q}e^{2V_{Q}}Q+T^{c}e^{-2V_{T}}\bar{T}^{c}+B^{c}e^{-2V_{B}}\bar{B}^{c})
(1-\Delta^{2}\theta^{2}\bar{\theta}^{2})\nonumber\\
&+& G\int d^{2}\theta d^{2}\bar{\theta}[(\bar{Q}~\bar{T}^{c})e^{2V_{Q}-2V_T}
(QT^{c})](1-2\Delta^{2}\theta^{2}\bar{\theta}^{2}+
\delta\bar{\theta}^{2}+\delta\theta^{2})~,
\label{Lsbhl}
\eea

where ${\cal L}_{YM}$ contains the usual SUSY kinetic terms for gauge fields,
$Q$ ($T^{c},B^c$) are SU(2) doublet (singlet) chiral quark
superfields. $\Delta$ and $\delta$ are SUSY soft breaking
parameters. Throughout this paper superfields will be denoted by 
capital letters and 
component fields by small letters except for the vectorfield
which is identifiable by its Dirac index. 
Reformulated in auxiliary fields eq.(\ref{Lsbhl}) corresponds to:

\bea
{\cal L}={\cal L}_{YM}&+&\int d^{2}\theta d^{2}\bar{\theta}
(\bar{Q}e^{2V_{Q}}Q+T^{c}e^{-2V_{T}}\bar{T}^{c}+B^{c}e^{-2V_{B}}\bar{B}^{c})
(1-\Delta^{2}\theta^{2}\bar{\theta}^{2})\nonumber\\
&+&\int d^{2}\theta d^{2}\bar{\theta}~\bar{H}_{1}
e^{2V_{H_{1}}}H_{1}(1-M_{H}^{2}\theta^{2}\bar{\theta}^{2})\nonumber\\
&-&\int d^2\theta\epsilon_{ij}(\mu_{0}H_{1}^{i}H_{2}^{j}(1+B_{0}\theta^{2})-
g_{T_{0}}H_{2}^{j}Q^{i}T^{c}(1+A_{0}\theta^{2}))\nonumber\\
&-&\int d^2\bar{\theta}\epsilon_{ij}(\mu_{0}\bar{H}_{1}^{i}\bar{H}_{2}^{j}
(1+B_{0}\bar{\theta}^{2})-g_{T_{0}}\bar{T}^{c}\bar{Q}^{i}\bar{H}^{j}_{2}
(1+A_{0}\bar{\theta}^{2}))  ~,
\label{Lauxsbhl}
\eea 

with $M_{H}^{2}=2\Delta^{2}+\delta^{2}$, $A_0-B_0=\delta$, 
$V_{H_1}=V_{Q}-V_T$ and $G=\frac{g_{T_{0}}^{2}}{\mu_{0}^{2}}$. It is an 
essential feature of propagating boundstates that they correspond to the
auxiliary fields of the binding effective coupling which receivekinetic
contributions at lower energies. Thus the auxiliary fields in
eq.(\ref{Lauxsbhl}) represent the two MSSM Higgs superfields.

A top mass is produced as a nontrivial solution of the following self 
consistent gap equation \cite{SBHL}:

\beq
G^{-1}=\frac{N_{c}\Delta^{2}}{16\pi^{2}}[(1+\frac{2m_{t}^{2}+\delta^{2}\alpha}
{2\Delta^{2}})ln(\frac{\Lambda^{4}}{(m_{t}^{2}+\Delta^{2})^{2}-m_{QQ^{c}}^{4}})
-\frac{2m_{t}^{2}}{\Delta^{2}}ln(\frac{\Lambda^{2}}{m_{t}^{2}})] ~,
\label{sgap}
\eeq
where $m^2_{QQ_c}$ is the self--consistent squark mass and
$\alpha=m^{2}_{QQ^{c}}/(\delta m_{t})$ is given by
\beq
\alpha^{-1}=1+\frac{GM_{H}^{2}N_{c}}{32\pi^{2}}ln(\frac{\Lambda^{4}}
{(m_{t}^{2}+\Delta^{2})^{2}-m_{QQ^{c}}^{4}}) ~.
\label{sgap2}
\eeq

One can see that, as consequence of the SUSY cancellation of quadratic 
divergences, the cutoff appears only logarithmically in this gap equation.
But the quadratic contributions of the right hand side of (\ref{sgap})
have to cancel the suppression factor of G to provide a nontrivial solution. 
Due to this fact, a nontrivial solution of the gap equation can be achieved 
only if the coupling $G$ is of the order $\frac{1}{\Delta^{2}}$ which is much 
larger than $\frac{1}{\Lambda^2}$.
One can of course introduce such couplings by hand but 
now the theory
lacks any explanation for the high scale of the boundstate breakup (the cutoff
scale). It seems to be advisable to investigate whether
an underlying theory is able to provide an explanation. 

%%%%%%%%%%%%%%%%%%%%%%%%%%%%%%%%%%%%%%%%%%%%%%%%%%%%%%%%%%%%%%%%%%%%%%%%%%

%\section{Heavy Field Exchange}

%%%%%%%%%%%%%%%%%%%%%%%%%%%%%%%%%%%%%%%%%%%%%%%%%%%%%%%%%%%%%%%%%%%%%%%%%%%%%

The most obvious idea for an underlying theory would be to 
consider a renormalizable theory with heavy degrees of freedom.
We will investigate whether it is possible to regain the four--scalar coupling
of (\ref{Lsbhl}) effectively by integrating out such heavy degrees of
freedom of an underlying theory. To get a reasonable structure for the 
effective theory it is necessary to assume a supersymmetric structure of
the heavy mass terms in the underlying theory. All physical components of
the heavy superfields acquire the same mass and have to be
integrated out.
Recently a way of integrating out heavy degrees of
freedom in superfield formalism has been used in SUSY GUTs 
\cite{dimpo}. But due to some subtleties concerning the supersymmetric
structure of our effective theory which will be discussed in an upcoming
paper we integrate out in component fields.

There are two different kinds of superfields that can be exchanged
in an underlying theory: Real (vector) superfields involving vectorfields
plus their fermionic superpartners and chiral superfields which,
to allow Yukawa couplings to quarks in agreement with R--parity,
should consist of scalars and their fermionic superpartners.

Following the usual approach of top--condensation we first have a look at
heavy gauge boson exchange. We introduce a new gauge group $G_S$,
a supersymmetric gauge coupling 
to the heavy gauge bosons $V_S$, the corresponding fieldstrength
contribution ${W_a}_S{W^a}_S$ and a coupling of $V_S$ to a heavy
Higgs sector to produce the mass of $V_S$. This Higgs sector should break
the symmetry $G_S$ in a supersymmetric way and should not couple to the
light sector. We assume a Higgs potential $P_{H_S}$ which is able to 
accomplish this
without looking into the problematic details. Additionally we will need
soft breaking terms (gaugino and heavy Higgs mass terms) denoted by
${\cal L}_{soft}$. 
Written in superfields the new interaction sector looks like:

\bea
{\cal L}_{NI} &=&  \int d^{2}\theta d^{2}\bar{\theta}
(\bar{Q}e^{g_SV_{S}}Q+T^{c}e^{g_SV_{S}}\bar{T}^{c}+
\bar{H}_{S}e^{g_SV_{S}}H_{S}\nonumber\\
&+& (\frac{1}{4}W_{sa}W_s^a + P_{H_S})\delta^2(\bar{\theta})+
h.c.+{\cal L}_{soft}~.   
\label{suv}
\eea

In component fields this corresponds to the following
Lagrangian:

\bea
{\it L}_{NI}&=&{\it L}_{{kin}_{pure}}(\tilde{q},\tilde{t},q,t,\lambda,\tilde{H})+
g_{s}\tilde{q}^{+}D_{s}\tilde{q}+F_{q}^{+}F_{q}+
\frac{1}{2}D_{s}D_{s}+M_{s}^{2}V_{\mu}V^{\mu}\nonumber\\
& &+(g_{s}\bar{q}~\bar{\sigma}^{\mu}V_{s\mu}q+
ig_{s}(\partial^{\mu}\tilde{q}^{+})V_{s\mu}\tilde{q}+
i\sqrt{2}g_{s}\tilde{q}^{+}\lambda_{s}q\\
& &+M_{s}\lambda_{s}\bar{\tilde{H}_{s}}+(q\leftrightarrow t)+h.c.)+...~,
\nonumber
\label{cl}
\eea

where ${\it L}_{{kin}_{pure}}$ denotes the pure, 
non--covariant kinetic terms and 
the dots contain the terms which will not contribute 
to order $1/M_S^2$ in the effective theory.
In order to integrate out the heavy degrees of freedom it is is 
important to notice that it will not be 
correct to just neglect the kinetic terms of
the heavy fields before extracting the equations of motion. We have
to deal with linear fermionic and quadratic bosonic  mass terms at the
same time. Additionally trilinear mass valued coupling
terms can appear in the Lagrangian (not now but in case of heavy 
scalar exchange), which decrease the order of
suppression of effective coupling terms. Thus derivative couplings coming from
heavy particle kinetic terms are not necessarily suppressed by orders
higher than $1/M_S^2$ and we have to take them into consideration. 
So we have to do the following: We  extract the equations of motion 
from the full
Lagrangian to get ``constituent relations'' for the heavy fields. These
relations
also include suppressed derivative terms of heavy fields
coming from their kinetic terms. We insert these relations to eliminate 
the heavy fields in lowest order. Then we reinsert the same relations
again to eliminate the suppressed derivative  terms of heavy fields. 
This gives us the correct effective theory up to order $1/M_S^2$.
We do the calculation in the Wess--Zumino gauge which can be done 
because all arguments are tree level arguments. 

\begin{figure}[ht]
\rotate[r]{
\epsfysize=14.5cm
\epsffile{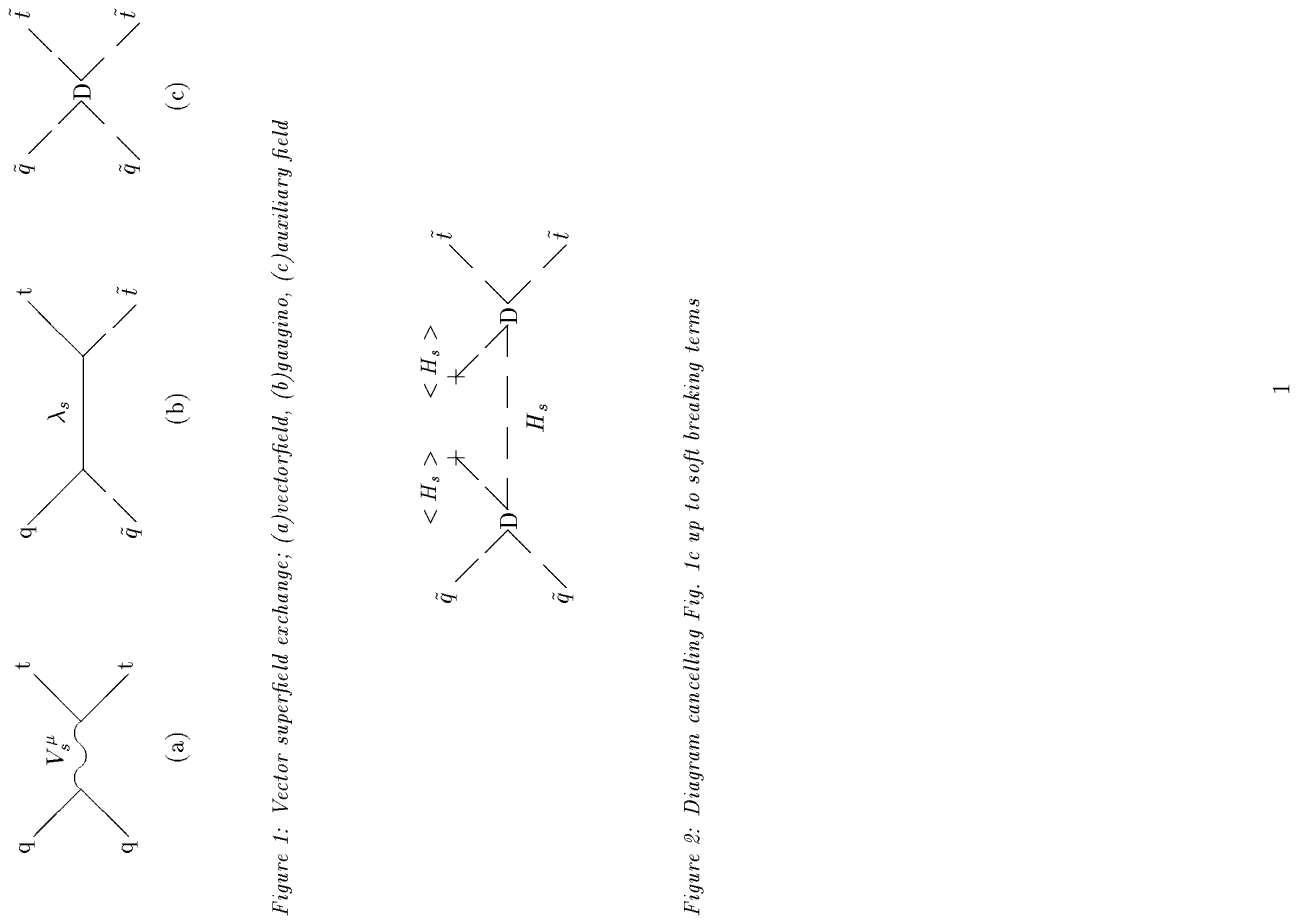}}
\end{figure}

The procedure outlined above leads to the following structure:
There are 3 phenomenons corresponding to
vector superfield exchange, which are gauge boson exchange, gaugino exchange,
and ``exchange'' of the auxiliary $D_{V_S}$--field (see Figure 1).
Each of them can roughly be connected to one 
four--superfield term:

(1): The vector boson exchange (Fig. 1a) produces component terms that fit
into the $1/M_S^2$--suppressed supersymmetric four--fermion term

\beq
{\cal L}_{eff1}=\frac{g_S^2}{M_{S}^2}
\int d^{2}\theta d^{2}\bar{\theta}[(\bar{Q}~\bar{T}^{c})
(QT^{c})]~.
\label{suveff1}
\eeq

However not all components of (\ref{suveff1}) are produced this way.
The remaining ones come from second order kinetic term contributions 
of gaugino exchange (b). 
Altogether all component terms corresponding to  (\ref{suveff1}) 
are correctly produced up to order $1/M_S^2$.

(2): The gaugino exchange (Fig. 1b) would not give any contribution to first order
in an exactly 
supersymmetric theory because the heavy VEV connects the two component
gaugino spinor to the Higgsino as its opposite chirality and the Higgsino
does not couple to the low energy sector. Only the soft breaking gaugino 
mass term gives soft breaking contributions of the order 
$(1/M_{S}) (\delta/M_{S})$ which can be written in
superfields as

\beq
{\cal L}_{eff2}=\frac{g_S^2}{M_{S}^2}
\int d^{2}\theta d^{2}\bar{\theta}[(\bar{Q}~\bar{T}^{c})
(QT^{c})](\delta\bar{\theta}^{2}+\delta\theta^{2})~.
\label{suveff2}
\eeq

with the soft breaking parameter $\delta$.
This happens the same way in SUSY GUTs~\cite{sugu1}.

(3): Also similar to SUSY GUTs is the situation of auxiliary 
field ``exchange'' (Fig. 1c) \cite{sugu2}: The contribution of Fig. 1c
is exactly cancelled
by the contribution of Fig. 2 in the supersymmetric case, 
only the soft breaking Higgs mass $\Delta^2$ leads to a soft breaking 
contribution of the order $(\Delta/M_{S})^2$:

\beq
{\cal L}_{eff3}=\frac{-g_S^2}{M_{S}^2}
\int d^{2}\theta d^{2}\bar{\theta}[(\bar{Q}~\bar{T}^{c})
(QT^{c})]2\Delta^{2}\theta^{2}\bar{\theta}^{2}~.
\label{suveff3}
\eeq

Finally we end up with an effective coupling term of the form:

\beq
{\cal L}_{eff}=\frac{g_S^2}{M_{S}^2}
\int d^{2}\theta d^{2}\bar{\theta}[(\bar{Q}~\bar{T}^{c})
(QT^{c})](1-2\Delta^{2}\theta^{2}\bar{\theta}^{2}+
\delta\bar{\theta}^{2}+\delta\theta^{2})~.
\label{suveff}
\eeq

The structure and the size of the soft breaking contributions of
the four superfield coupling in equation (\ref{Lsbhl}) are correctly 
regained but,
not surprisingly, the suppression factor of the four--superfield 
coupling is of the
order of the heavy mass scale while we want it to be of the soft 
breaking scale.

Still this is not the whole story: For a realistic model it 
is necessary to take into consideration also the electroweak gauge couplings,
which will lead to even more fundamental
problems of heavy vector superfield exchange.
We write the kinetic terms of the full theory as:

\bea
{\cal L}_{kin} &=& \int d^{2}\theta d^{2}\bar{\theta}
(\bar{Q}e^{(g_SV_{S}+g_2V_2+g_{Q}V_1)}Q+
T^{c}e^{(g_SV_{S}+g_{T}V_1)}\bar{T}^{c} \nonumber\\
&+& \bar{H}_{S}e^{g_SV_{S}+g_{H_S}V_1}H_{S}
+\frac{1}{4}(W_{sa}W_s^a + {W_{ew}}_aW_{ew}^a)\delta^2(\bar{\theta})+h.c.
\label{givv}
\eea

Now we have a look at the gauge structure: In the effective
theory we are confronted
with different covariant derivations 
$D^{\mu}_Q={\partial}^{\mu}+ig_2V_2^{\mu}+ig_QV^{\mu}_1$
and $D_T^{\mu}={\partial}^{\mu}+ig_TV_1^{\mu}$ for $q, 
\tilde{q}$ and $t, 
\tilde{t}$ 
respectively. One of the
dimension 6 effective coupling terms is the
four--scalar coupling term which
includes two covariant derivatives. As described in above in the framework
of auxiliary fields,  
this term has to be identified with
the kinetic term of the scalar boundstate $h_1={\tilde{q}}^{\dagger}\tilde{t}$
at the cutoff scale. Therefore it should have the structure
$[D^{\mu}_{QT}(\tilde{q}^{\dagger}\tilde{t})]^{\dagger}
D^{\mu}_{QT}(\tilde{q}^{\dagger}\tilde{t})$ with $D^{\mu}_{QT}=
{\partial}^{\mu}+ig_2V_2^{\mu}+i(g_Q-g_T)V^{\mu}_1$.   
But the term generated by integrating out 
a heavy vector field is the $D_QD_T$ term 
$[\tilde{q}^{\dagger}D_{Q}\tilde{q}]^{\dagger}
\tilde{t}^{\dagger}D_{T}\tilde{t}$. (One covariant derivative of the
$\tilde{q}$ and $\tilde{t}$  scalar kinetic term is used up by the 
intermediating heavy gauge Boson.) 
This is, of course, a gauge invariant coupling term as well, but it is
simply the wrong one. It does not fit into the structure of eq.(\ref{Lsbhl})
and it is not consistent with low energy propagating boundstates whose 
structure should resemble the structure of the leading order binding 
coupling terms. 
Heavy vector superfield exchange is therefore not able to give
the correct gauge structure for finetuned 
supersymmetric dynamical electroweak breaking.

There seems to occur another problem connected to the phenomenon that 
the effective dimension 6 operators do not show the correct 
supersymmetric structure. An investigation of this point is in progress.

The alternative to vector superfield exchange would be the exchange of 
massive chiral superfields which do not break any gauge symmetry. 
In this case
the exchanged chiral superfields have to resemble exactly the gauge
structure of the auxiliary fields describing the effective couplings.
Therefore problems with a wrong gauge structure like in vector
superfield exchange are safely excluded.
A point we made in the beginning of this discussion can be seen
quite impressively now: The part of the underlying theory responsible 
for the effective four--coupling 
includes only F--terms (Yukawa coupling term and the  $\mu$--term) 
and the kinetic terms of the heavy fields. 
The F--terms which do not contain any derivatives 
in component fields will never be able to lead to the D--term in our 
effective theory (simply because the derivatives that occur there 
can not emerge
out of nothing.) So it is a crucial point to take into consideration the 
kinetic terms of the heavy fields which produce  the effective
D--term in a perfectly supersymmetric way.
If we neglect the Yukawa term for the second Higgs field the effective
F--term is zero and the structure of the effective theory is in agreement 
with (\ref{Lsbhl}) except for the notorious scale problem for $G$.
Once again the big suppression factor of effective coupling  $G$
does not allow dynamical symmetry breaking.
\footnote{However we are inclined to doubt the physical meaning
of a concept of producing effective scalar boundstates by heavy scalar 
exchange in general. A discussion of this point is in preparation.}.

%%%%%%%%%%%%%%%%%%%%%%%%%%%%%%%%%%%%%%%%%%%%%%%%%%%%%%%%%%%%%%%%%%%%%%%%%%

%\section{Connecting Dynamical Electroweak Breaking to SUGRA}

%%%%%%%%%%%%%%%%%%%%%%%%%%%%%%%%%%%%%%%%%%%%%%%%%%%%%%%%%%%%%%%%%%%%%%%%%%%%%

From this discussion we conclude that a renormalizable underlying 
gauge field theory is not
able to produce propagating effective Higgs fields for several reasons.
One has to look therefore for alternative concepts.

An interesting approach to gain a low suppression scale for $G$
has been suggested by Ellwanger
\cite{ELLW} in the framework of supersymmetric nonlinear sigma models.
However this remains an effective theory approach.
It is not clear to us how it could be connected to a fundamental 
theory behind.

A good way to understand the problem of separating the scales  $\Delta$
and $\Lambda$
is to have a careful look at the effective Lagrangians. It is a fundamental
principle of top condensation that the SM serves as a good effective
theory below the cutoff scale  $\Lambda$ which means that the SM 
Lagrangian can be
identified with the top condensation Lagrangian at the cutoff scale.
This requirement is best formulated by rewriting the top condensation 
Lagrangian in auxiliary field formalism and 
corresponds to conditions for the 
SM parameters at $\Lambda$, the 3 so called constituent conditions 
\cite{BHL}:
At the cutoff scale
the top Yukawa coupling $g_t$  and the four Higgs coupling $\lambda$ 
of the SM have
to run into a pole and, which is important for our discussion,
the SM Higgs mass parameter $m^2$ has to be identified with the inverse
four--fermion coupling $-G^{-1}$ of top condensation. While the first two
conditions have to be fulfilled by renormalization group running of
$g_t$ and $\lambda$ and predict certain low energy values for these
parameters, the situation for $m^2$ is different. The running
of $m^2$ from the electroweak scale up to the cutoff has to change its
sign and shift its scale from electroweak to cutoff size.
This running behaviour cannot stem from mere
renormalization group running. The SM renormalization
group running of the parameter $m^2$ cannot change its sign, i. e. 
the SM does not allow radiative breaking of the electroweak symmetry. 
In contrast with renormalization group running,
the kind of running that is responsible for the identification
of $m^2$ with $-G^{-1}$ at the cutoff scale in top condensation
has to include quadratic contributions.  This phenomenon is connected 
to the fact, that the whole concept of Nambu--Jona-Lasinio like dynamical
symmetry breaking is  based on the relevance of quadratic
divergences in the self consistent gap equation. The quadratic running
behaviour of $m^2$ can be more or less 
understood as the effective temperature running of $m^2$ which
comes up to a mass contribution of the order temperature. 

In case of SUSY top condensation the situation is different:
The parameter that has to be identified with $G$ at the cutoff scale
is $\frac{g_t^2}{\mu^2}$ involving the SUSY invariant $\mu$--term. 
This $\mu$--term 
necessarily produces a positive scalar mass term $\mu^2$ after integrating 
out F--terms. Therefore no change of sign has to take place to identify 
$\frac{g_t^2}{\mu^2}$ with $G$. 
Now as the contribution
of temperature breaks SUSY \cite{lal} it must have the structure
of softbreaking scalar mass terms and cannot show the structure of the 
supersymmetric $\mu$--term. Thus the $\mu$--term is not touched 
by temperature running, contrary to standard top
condensation the identification with G has to be achieved by mere 
renormalization group running. Notice that this fact is essential for making
a small $G$--coupling suppression factor possible. Otherwise the suppression
factor would necessarily stay at the cutoff scale due to temperature
arguments.

Now one should have a look at the renormalization group running
of $\frac{g_t^2}{\mu^2}$. Both $g_t^2$ and $\mu^2$ run into a pole 
driven by the same contributions. These pole making contributions
cancel\footnote{One cannot see this cancellation by expanding in the
coupling constant, as this expansion obviously breaks down in the pole
region. However the cancellation is correctly produced by the first order
$1/N_c$ $\beta$--functions.}  \cite{SBHL}, the rest is logarithmic 
running that does not change the order
of the scale drastically. Noticing that low energy $g_t^2$ is approximately 1,
we see that the low energy value of $\mu^2$ indeed gives the order of the
$G$--coupling suppression factor. This makes quite an interesting difference
compared to standard top condensation.

The conclusion of this discussion is that the problem how to separate  
softbreaking
and cutoff scale is directly related to the SUSY
$\mu$--problem \cite{DUDAS}. This $\mu$--problem can be stated as follows:
The $\mu$--term in a MSSM has to be of the order softbreaking scale to
allow electroweak symmetry breaking. But being a supersymmetric term
it is not restricted by a SUSY breaking scale and should
therefore be expected to be naturally of a high (e. g. the Planck)
scale. Here we find again our problem of separating the scales.
As we assume the MSSM to be an effective theory of SUSY top condensation,
the two problems are identical for us, just formulated once in the
framework of an effective theory (MSSM) and the other time in the framework
of the underlying theory (top condensation).
Now there exist solutions of the $\mu$--problem in connection with
supergravity.
The next step should be to see whether these solutions
can be maintained in the framework of a constituent Higgs model. 

The basic idea of these solutions of the $\mu$--problem is to forbid the direct
$\mu$--term in the Lagrangian (something which could be natural in a
superstring scenario) and to regain a low energy $\mu$--term from other
sources within supergravity \cite{mu}. A quite general approach is to couple 
the Higgses 
to some singlet from the hidden sector like: 
$$\int d^{2}\theta d^{2}\bar{\theta}\frac{1}{M_{P}} Z^{+}H_{1}H_{2}\ + h.c.$$
In  global SUSY this term can be written as an F--term:
\beq
\int d^{2}\theta d^{2}\bar{\theta}\frac{1}{M_{P}} Z^{+}H_{1}H_{2}=-\frac{1}{4}
\int d^{2}\theta \frac{\bar{D}^2Z^{+}}{M_{P}}H_{1}H_{2}
\label{fmu}
\eeq
If SUSY is broken in the hidden sector by a vacuum expectation value of the 
auxiliary field of Z at a scale $<F_{Z}>\sim (10^{10}-10^{11})^{2}$, 
an effective $\mu$--term of the desired scale 
$\mu = \frac{<F_{Z}>}{M_P} \sim 10^4 GeV$ is produced. 

Now, to apply this procedure to a constituent Higgs model, it is
necessary to be able to interpret the Higgs fields of (\ref{fmu}) as
auxiliary fields. As above we use the condition that the kinetic term of 
$H_{2}$ vanishes at the cutoff scale $\Lambda$ and write the Lagrangian 
%of the Higgs sector:
%\bea
%{\cal L}_{H}=\int d^{2}\theta d^{2}\bar{\theta}\bar{H_{1}}exp(2V_{H_{1}})H_{1}
%(1-M_{H}^{2}\theta^{2}\bar{\theta}^{2})

\bea
{\cal L}={\cal L}_{YM}&+&\int d^{2}\theta d^{2}\bar{\theta}
(\bar{Q}e^{2V_{Q}}Q+T^{c}e^{-2V_{T}}\bar{T}^{c}+B^{c}e^{-2V_{B}}\bar{B}^{c})
(1-\Delta^{2}\theta^{2}\bar{\theta}^{2})\nonumber\\
&+&\int d^{2}\theta d^{2}\bar{\theta}~\bar{H}_{1}
exp(2V_{H_{1}})H_{1}(1-M_{H}^{2}\theta^{2}\bar{\theta}^{2})\nonumber\\
&-&\int d\theta^{2}\epsilon_{ij}(\frac{\bar{D}^2Z^{+}}{M_{P}}H_{1}^{i}H_{2}^{j}
(1+B_{0}\theta^{2})-
g_{T_{0}}H_{2}^{j}Q^{i}T^{c}(1+A_{0}\theta^{2}))\nonumber\\
&-&\int d\bar{\theta}^{2}\epsilon_{ij}(\frac{D^2Z}{M_{P}}\bar{H}_{1}^{i}
\bar{H}_{2}^{j}
(1+B_{0}\bar{\theta}^{2})-g_{T_{0}}\bar{T}^{c}\bar{Q}^{i}\bar{H}^{j}_{2}
(1+A_{0}\bar{\theta}^{2}))~.
\label{Lauxnew}
\eea 

The Euler--Lagrange equations give the following constituent relations for the
fields $H_{1}$, $H_{2}$:
\beq
\frac{\bar{D}^2Z^{+}}{M_{P}}H_{1}(1+B_{0}\theta^{2})=g_{T_{0}} 
QT^{c}(1+A_{0}\theta^{2})~,
\label{com1}
\eeq
\beq
\frac{Z^{+}}{M_{P}}H_{2}(1+B_{0}\theta^{2})=-\frac{1}{4}\bar{H}_{1}
e^{2V_{H_{1}}}(1-M_{H}^{2}\theta^{2}\bar{\theta}^{2})~.
\label{com2}
\eeq

At low energies the couplings to the hidden sector field Z are suppressed
by the Planck scale and therefore neglectable with one exception: the
coupling to the VEV of the auxiliary field $F_Z$ is important. Like in
the fundamental Higgs case the VEV produces a  mass
term $\mu_{0}=\frac{<F_Z>}{M_P}$ of the order $10^4$ GeV. 
Inserting $<F_Z>$ into (\ref{com1}) and (\ref{com2}) and neglecting all terms
suppressed by $M_P$  we get the 
following constituent relations in the broken phase: 

\beq
H_{1}(1+B_{0}\theta^{2})=\frac{g_{T_{0}}}{\mu_0}QT^{c}(1+A_{0}\theta^{2})~,
\label{co1}
\eeq
\beq
H_{2}(1+B_{0}\theta^{2})=
\frac{-\bar{D}^2}{4\mu_0}\bar{H}_{1}
e^{2V_{H_{1}}}(1-M_{H}^{2}\theta^{2}\bar{\theta}^{2})~,
\label{co2}
\eeq

which are exactly the constituent relations of Bardeen et al. \cite{SBHL}.
Thus we get a reasonable four--fermion theory in the SUSY broken phase
which becomes nonlocal at the cutoff scale $\Lambda$ not due to a
propagating heavy gauge boson but due to some underlying finite theory
at the Planck scale, e. g. a string theory. Consequently the scale  
$\Lambda$ has to be identified with the Planck scale.

The situation in the unbroken phase is much less attractive. The
constituent relations (\ref{com1}), (\ref{com2}) are highly nonlinear, 
it seems to be impossible to eliminate the auxiliary fields. Still
the existence of this non--dynamical relation means that there is no 
independent Higgs degree of freedom in the theory. We claim that this
nonlinear realization is the only possibility to achieve a critical
Nambu--Jona-Lasinio like gap equation in a supersymmetric framework.
These are once more the summarized arguments leading to this statement:

We want to produce a Higgs bound state that propagates up to a scale
$\Lambda$ higher than the scale where SUSY becomes relevant, i.e. 
the SUSY soft breaking scale.
Because of the cancellation of quadratic divergences a critical
Nambu--Jona-Lasinio gap equation in a SUSY theory requires a four--fermion
coupling G suppressed only by the soft breaking scale.
A renormalizable underlying gauge field theory, i.e. an exchange of heavy 
particles with mass of the cutoff scale $\Lambda$
would not be able to separate the scales of  $\Lambda^2$ and $G^{-1}$.
The alternative is to maintain the nonrenormalizable structure
up to the Planck scale. As the constituent conditions
for RG running parameters 
in a SUSY theory of propagating boundstates are formulated without involving
temperature running, the SUSY $\mu$--term gives the scale for the
suppression factor of G.
Since the SUSY theory with Higgses has to be an effective theory of the
Nambu--Jona-Lasinio model, the structure that produces the low
suppression of G has to appear as a solution to the $\mu$--problem
in this effective SUSY theory.
But the known solutions to the $\mu$--problem correspond to a nonlinear
constituent relation in a constituent Higgs model.

While the nonlinear constituent relation in the 
unbroken phase is not a highly attractive feature of this model,
the model shows apart from this obstacle a number of nice 
features: 

The model gives a natural solution to the $\mu$--problem.
There is no need of a symmetry like R--symmetry or Peccei--Quinn symmetry 
that forbids the $\mu$--term in the superpotential like 
in a theory with fundamental Higgses.
An effective $\mu$--term in the superpotential would be produced by
a 4--superfield D--term
$$\int d^{2}\theta d^{2}\bar{\theta}\frac{1}{M_P^2}\bar{Q}~\bar{T}^cQT^c~,$$ 
which is suppressed by the Planck--scale. But adding
this additional interaction term to the Lagrangian (\ref{Lauxnew}) 
only gives highly
suppressed additional contributions to the low energy theory. The low
energy four fermion interaction becomes:
$$\int d^{2}\theta d^{2}\bar{\theta}\frac{1}{M_P^2}\bar{Q}~\bar{T}^cQT^c
+\int d^{2}\theta d^{2}\bar{\theta}\frac{1}{\Delta^2}\bar{Q}~\bar{T}^cQT^c
\simeq \int d^{2}\theta d^{2}\bar{\theta}\frac{1}{\Delta^2(1-\frac{\Delta^2}{M_
P^2})}
\bar{Q}~\bar{T}^cQT^c~. $$ 
By 
constructing the $\mu$--term from four--superfield interactions the low 
scale becomes
the dominating one. There is no necessity to forbid any type of
possible interaction.

Another interesting feature is the fact that the argument of Hasenfratz 
et al. \cite{has} lanced against four--fermion top condensation is not 
valid in our case. This argument states that higher dimensional 
operators can change the predictions of a four fermion top 
condensation model arbitrarily so that this model
would turn out to be just a different parametrization of the SM. 
In our case higher dimensional operators are suppressed by higher orders
of the Planck scale while the suppression factor of the effective
four--fermion coupling is just the SUSY softbreaking scale $\Delta$. Thus 
these higher order operators are irrelevant in low energy physics.

%\section{Conclusions}

In summary we have analyzed the possibilities for theories underlying
supersymmetric Nambu--Jona-Lasinio like models 
which produce a supersymmetric propagating Higgs bound state
(e.g. supersymmetric top condensation).  
The general problem of such a concept is to find
some mechanism which is able to separate the 
suppression scale of the effective four--superfield coupling from
the effective cutoff scale, the scale of new physics in the 
underlying theory. It turns out that a renormalizable 
gauge field theory, which is a theory of heavy particle exchange,
is not able to achieve this. The approach of underlying 
vector superfield exchange additionally is
not able to produce the gauge structure of the four--superfield coupling
that would be needed to build Higgs boundstates. 
We argued that the only reasonable alternative in the framework
of established physical concepts is to couple dynamical
electroweak breaking to supergravity. In this scenario 
the correct effective structure of a
Nambu--Jona-Lasinio like model is produced as a result of SUSY breaking in
the hidden sector. Such a concept 
automatically implies the desired vanishing of the $\mu$--term 
in the SUGRA superpotential
and suppresses unwanted higher dimensional operators that could
change the predictions of top condensation. 
A drawback of this approach is the fact 
that, in spite of being a model without independent Higgs 
degrees of freedom,
it does not allow to integrate out the auxiliary fields linearly.

\vspace{.5cm}
{\bf Acknowledgments:} We would like to thank E. Dudas who called 
our attention at the relevance of the $\mu$--problem in this context and
A. Blumhofer, M. Lindner, U. Nierste and  M.~Yamaguchi
for useful discussions and helpful comments
on the draft version of this paper.

This work is in part supported by DFG (contract Li 519/2-1).

%%%%%%%%%%%%%%%%%%%%%%%%%%%%%%%%%%%%%%%%%%%%%%%%%%%%%%%%%%%%%%%%%%%%%%%%%
%%%%                           References                            %%%%
%%%%%%%%%%%%%%%%%%%%%%%%%%%%%%%%%%%%%%%%%%%%%%%%%%%%%%%%%%%%%%%%%%%%%%%%%
%\newpage

%%%%%%%%%%%%%%%%%%%%%%%%%%%%%%%%%%%%%%%%%%%%%%%%%%%%%%%%%%%%%%%%%%%%%%%%%
\end{document}